\documentclass{emulateapj}

\usepackage{amsmath}
\usepackage{natbib}
\usepackage{graphicx}
\usepackage{color}
\usepackage[breaklinks,colorlinks,citecolor=blue]{hyperref}
\def\nn{\nonumber}
\def\st{\sin\theta}
\def\ct{\cos\theta}
\def\sst{\sin^2\theta}
\def\cct{\cos^2\theta}
\def\be{\begin{equation}}
\def\ee{\end{equation}}
\def\ben{\begin{eqnarray}}
\def\een{\end{eqnarray}}
\def\WH{\Omega_{\rm H}}
\def\R2{R_2}
\def\C{C^{(\frac{3}{2})}}

\begin{document}
\title{Analytic properties of force-free jets in the Kerr spacetime -- I}

\slugcomment{Submitted to \apj}

\author{
  Zhen Pan,\altaffilmark{1}
  Cong Yu,\altaffilmark{2,3}
 }

\altaffiltext{1}{Department of Physics, University of California,
One Shields Avenue, Davis, CA, 95616, USA,  {\tt zhpan@ucdavis.edu}}
\altaffiltext{2}{Yunnan Observatories, Chinese Academy of Sciences, Kunming 650011, China, {\tt cyu@ynao.ac.cn}}
\altaffiltext{3}{Key Laboratory for the Structure and Evolution of Celestial Objects, Chinese Academy of Sciences, Kunming 650011, China}

\shorttitle{Analytic properties of force-free jets in the Kerr spacetime}
\shortauthors{Z. Pan and C.  Yu}

\begin{abstract}
Blandford-Znajek (BZ) mechanism describes a process extracting
rotation energy from a spinning black hole (BH) via magnetic field
lines penetrating the event horizon of central BH. In this paper,
we present a perturbation approach to study force-free
jets launched by the BZ mechanism, and its two immediate
applications: (1) we present a high-order split monopole
perturbation solution to the BZ mechanism, which accurately pins
down the energy extraction rate $\dot E$ and well describes the
structure of BH magnetosphere for all range of BH spins ($0\leq
a\leq 1$); (2) the approach yields an exact constraint for the
monopole field configuration in the Kerr spacetime, $I = \Omega
(1-A_\phi^2)$, where $A_\phi$ is the $\phi-$component of the vector
potential of electromagnetic field, $\Omega$ is the angular velocity
of magnetic field lines and $I$ is the poloidal electric current.
The constraint is of particular importance to benchmark the
accuracy of numerical simulations.
\end{abstract}

\keywords{gravitation -- magnetic fields -- magnetohydrodynamics}

\bigskip\bigskip

\section{Introduction}
The Blandford-Znajek mechanism
\citep{Blandford1977d} is believed to be an efficient way to
extract rotation energy out of a spinning BH, which operates in BH
systems on all mass scales, from stellar-mass BHs of gamma ray
bursts to supermassive BHs of active galactic nuclei. In the past
decade, general relativistic magnetohydrodynamic simulations have
shown a physical picture of jets launched by the BZ mechanism
\citep{Komissarov2001,Komissarov2004e,Komissarov2004,Komissarov2005,
Komissarov2007d, McKinney2004f,  McKinney2005,2008MNRAS.388..551T,
Tchekhovskoy2010,  2011MNRAS.418L..79T, 2012MNRAS.423L..55T,
Penna2013a, 2013Sci...339...49M}, which shows that purely
poloidal magnetic fields, $\mathbf{B_P}$, appear around a
non-rotating BH, while a spinning BH induces poloidal electric
field $\mathbf{E_P}$ and toroidal magnetic field $\mathbf{B_T}$,
thus an outward Poynting flux $\mathbf{E_P\times B_T}$ is
generated, and rotation energy of the spinning BH is extracted in
the form of Poynting flux.

Independent of time-dependent simulations, \citet{2013ApJ...765..113C, 2014ApJ...788..186N} numerically solved the general relativistic Grad-Shafranov (GS) equation which governs the structure of the stationary force-free magnetosphere around a Kerr BH. It is known that the GS equation is a second-order differential equation of $A_\phi$ depending on the magnetic field angular velocity $\Omega$ and the poloidal electric current $I$.
The GS equation was shown to be an eigenvalue problem with eigenfunction $A_\phi$ and eigenvalues $I$ and $\Omega$ to be determined. To solve the GS equation self-consistently,  eigenvalues $I$ and $\Omega$ were adjusted to make magnetic fields smoothly cross {\it two} light surfaces. In this way, they confirmed the results of previous simulations for asymptotically monopole and paraboloidal magnetic field. But for the asymptotically uniform magnetic field, there is only {\it one} light surface, which is not sufficient for determining the {\it two} eigenvalues $I$ and $\Omega$, so they argued that there are infinitely many solutions for the asymptotically uniform field \citep{2013ApJ...765..113C, 2014ApJ...788..186N, 2015arXiv150306788Y}. While time-dependent simulations conducted by different groups seem to converge to similar solutions (e.g., \cite{Komissarov2005, Komissarov2007d, Alic2012}).  But different simulations did not precisely match, especially in the range of extreme spins, and there is no way to tell which simulation is more robust.

To settle down all these uncertainties and controversies, independent analytic works are of great value \citep{Tanabe2008, Pan2014, Pan2015}. Though \cite{2013ApJ...765..113C} criticized that the analytic perturbation approach did not appreciate the critical role of light surfaces, it is no coincidence that previous perturbation solutions are consistent with simulations, especially for non-extremal Kerr BHs (e.g. \cite{McKinney2004f, Tchekhovskoy2010}). In fact, perturbation solutions obtained are analytical and smooth, which certainly satisfy the requirement of magnetic field lines smoothly crossing the light surfaces. So perturbation approach is consistent with the eigenvalue approach proposed by \citet{2013ApJ...765..113C}, and the only difference is that the eigenvalue approach implements the smoothly crossing requirement explicitly, while the perturbation approach implements the requirement implicitly. 

In this paper, we propose a perturbation approach to investigate the BZ mechanism analytically, which in principle enables us to obtain perturbation solutions accurate to any order of the BH spin. Following this approach, we present a high-order monopole perturbation solution to the BZ mechanism, and derive an exact constraint relation for monopole magnetic field, which provides a criteria for testing the accuracy of numerical simulations.
The paper is organized as follows. Basic equations governing  stationary axisymmetric force-free magnetosphere in the Kerr spacetime are summarized in Section \ref{sec: basic}. In Section \ref{sec: perturbation}, we present our perturbation method and apply it to the monopole magnetic field. We present results in Section \ref{sec: results} and discussion in Section \ref{sec: discussions}. We conclude in Section \ref{sec: conclusions}.

\section{Basic equations}
\label{sec: basic}

Stationary axisymmetric force-free
electromagnetic fields in the Kerr spacetime are determined by
three functions $A_\phi$, $\Omega(A_\phi)$, $I(A_\phi)$.
Non-trivial components of Faraday tensor $F_{\mu\nu}\equiv
\partial_\mu A_\nu-\partial_\nu A_\mu$ in the
Kerr-Schild coordinate \citep{McKinney2004f, 1963PhRvL..11..237K}
could be expressed as follows
\citep{Blandford1977d,McKinney2004f,Pan2014} \be F_{r\phi} =
-F_{\phi r}=A_{\phi,r} \ , F_{\theta\phi} = - F_{\phi\theta} =
A_{\phi,\theta} \ , \ee \be F_{tr} = -F_{rt} = \Omega A_{\phi,r} \
, F_{t\theta}= - F_{\theta t} = \Omega A_{\phi,\theta} \ , \ee \be
F_{r\theta} = -F_{\theta r} = \sqrt{-g}B^{\phi} \ , \ee where \be
B^\phi = - \frac{I \Sigma + (2 \Omega r - a) \sin\theta
A_{\phi,\theta}} {\Delta \Sigma \sin^2\theta} \ , \label{eq:Bphi}
\ee and $\Sigma=r^2+a^2\cos^2\theta$, $\Delta=r^2-2r+a^2$,
$\sqrt{-g}=\Sigma\sin\theta$. The energy conservation equation
reads \ben\label{eq:GS} &-&\Omega \left[(\sqrt{-g}F^{tr})_{,r} +
(\sqrt{-g}F^{t\theta})_{,\theta} \right] + F_{r\theta}I'(A_\phi) \nn\\
&&+ \left[(\sqrt{-g}F^{\phi r})_{,r} +
(\sqrt{-g}F^{\phi\theta})_{,\theta} \right] = 0 \ , \een which is
also known as the Grad-Shafranov (GS) equation.

In the Schwarzschild spacetime, $I = \Omega = 0$, the GS equation is
simplified as \be \mathcal L A_\phi =0, \ee where the operator \be
\mathcal L\equiv\frac{1}{\sin\theta}\frac{\partial}{\partial
r}\left(1-\frac{2}{r}\right)\frac{\partial}{\partial r}
 +\frac{1}{r^2}\frac{\partial}{\partial \theta}\frac{1}{\sin \theta}\frac{\partial}{\partial \theta} .
\ee The Green's function $G(r,\theta; r_0, \theta_0)$ to the
operator $\mathcal L$ defined by \be \mathcal LG(r,\theta; r_0,
\theta_0) = \delta(r-r_0) \delta(\theta-\theta_0), \ee is
available \citep{Blandford1977d, 1974PhRvD..10.3166P}. Note that
the solution satisfies the boundary conditions that
$G(r,\theta; r_0, \theta_0)$ is finite at $r=2$ and approaches to
zero at infinity.

\section{Perturbation method}
\label{sec: perturbation}

It is notoriously difficult to exactly solve the GS equation (\ref{eq:GS}) due to its nonlinearity. We adopt the perturbation technique to attack this problem. \citet{Blandford1977d} first put forward the monopole perturbation solution up to order of $O(a^2)$ and \citet{Pan2015} extended the solution to $O(a^4)$. In this paper, we adopt a generalized perturbation approach, which, in principle, enables us to obtain perturbation solutions to any high order. We find that truncation at the eighth-order is good enough to precisely match current state-of-the-art numerical simulations. This method can be applied to other type of field configuration as well, though tedious calculations are inevitable. For simplicity, we focus on the split-monopole field throughout this paper.

For monopole magnetic field, the Schwarzschild metric solution to
the GS equation (\ref{eq:GS}) writes as
\be
\Omega_0 = 0, \quad I_0 = 0, \quad A_\phi = A_0 = -\ct.
\ee
For corresponding Kerr metric solution, we define $\omega = \Omega(A_\phi)|_{r\rightarrow\infty}$, $i =  I(A_\phi)|_{r\rightarrow\infty}$, and expand them in series,
\ben
A_\phi &=& A_0 + a^2 A_2 + a^4 A_4 +  O(a^6), \nn\\
\omega &=& a \omega_1 + a^3\omega_3 + a^5\omega_5 + O(a^7),\nn\\
i &=& a i_1 + a^3 i_3 + a^5i_5+ O(a^7).
\een
As done by \citet{Pan2015}, $\Omega$ and $I$ could be expressed in terms of $\omega$ and $i$ respectively.
With the above notations, the GS equation (\ref{eq:GS}) could be decomposed as a set of linear equations
\be
\mathcal L A_n(r,\theta) = S_n(r,\theta; i_{n-1},\omega_{n-1}) \quad (n=2,4,6,...) .
\ee
These equations could be solved one by one as follows.

Accurate to $O(a^2)$, GS equation (\ref{eq:GS}) is written as \be
\mathcal L A_2(r,\theta) = S_2(r, \theta; i_1, \omega_1). \ee
Before solving the differential equation, $i_1$ and $\omega_1$
need to be specified by two constraints. The Znajek regularity
condition \citep{Znajek1977b} which requires $B^\phi$ to be finite on horizon gives
\be i_1 =
\sst\left(\frac{1}{4}-\omega_1\right), \ee and the convergence
constraint that perturbation solution $A_2(r,\theta)$ should be
convergent from horizon to infinity gives (see \citet{Pan2015} for
details) \be i_1 = \omega_1 \sst. \ee Consequently, \be i_1 =
\omega_1 \sst, \quad \omega_1 = \frac{1}{8}. \ee With $i_1$ and
$\omega_1$ specified, the source function $S_2(r,\theta)$ is
determined, so we can write the solution $A_2(r,\theta)$ as an
integral of the Green's function \be A_2(r,\theta) = \int_2^\infty
dr_0 \int_0^\pi d\theta_0 S_2(r_0,\theta_0) G(r,\theta; r_0,
\theta_0). \ee In this way, we work out all the three functions,
$A_\phi, I(A_\phi)$ and $\Omega(A_\phi)$ up to $O(a^2)$. Refer to
\citet{Blandford1977d}  for explicit form of $A_2(r,\theta)$, where
on inner/outer boundary \be A_2(2, \theta) = R \sst\ct, \quad
\lim_{r\rightarrow\infty} A_2(r, \theta) = 0, \ee
with $R= (6\pi^2-49)/72$.\\

We could obtain solutions up to $O(a^4)$  in a
similar manner \citep{Pan2015}. The Znajek horizon condition and the convergence
constraint give \be i_3 = \omega_3 \sst, \quad
\omega_3=\frac{1-4R}{64}\sin^2\theta + \frac{1}{32}, \ee and the
integral of Green's function  gives \ben
&&A_4(2,\theta) = p \sst \C_1(\ct) + q \sst \C_3(\ct), \nn\\
&&\lim_{r\rightarrow\infty} A_4(r, \theta) = 0.
\een
where
\ben
\C_1(\ct) &=& 3\ct, \nn\\
\C_3(\ct) &=& -\frac{15}{2}\ct + \frac{35}{2}\cos^3\theta,
\een
and
\ben
p &=&-\frac{9 \sigma }{14}+\frac{13\zeta(3)}{3920}-\frac{\pi ^2 \left(19385 + 774 \pi^2\right)}{181440}+\frac{17929399}{7620480}\nn\\
&\simeq& 1.7\times 10^{-2},\nn\\
q&=&-\frac{6 \sigma }{7}+\frac{321 \zeta (3)}{9800}-\frac{\pi ^2 \left(48955+ 918 \pi^2\right)}{201600} +\frac{2012505017}{508032000}\nn\\
&\simeq& 9.0\times 10^{-4},
\een
with
\ben
\sigma
&=&\Re \int_2^{\infty } \frac{2 \text{Li}_2\left(\frac{r}{2}\right) \log \left(\frac{r}{2}\right)}{\left(1-\frac{2}{r}\right) r^2} \, dr = 1.3529...\nn\\
\zeta(3)
&=&\frac{1}{2}\int_0^\infty \frac{t^2}{e^t-1} dt = 1.2020...
\een\\
where $\Re$ designates the real part and $\text{Li}_2(x)$
is the second-order Jonquiere's function.

Take one step further, we could obtain
quantities of $O(a^5)$ \be i_5 =\omega_5\sst, \quad \omega_5
=\frac{1}{64}(1-0.037 \sin^2\theta + 0.155 \sin^4\theta), \ee
using the Znajek horizon condition and the convergence constraint.

\section{Results}
\label{sec: results}

\begin{figure}
\includegraphics[scale=0.42]{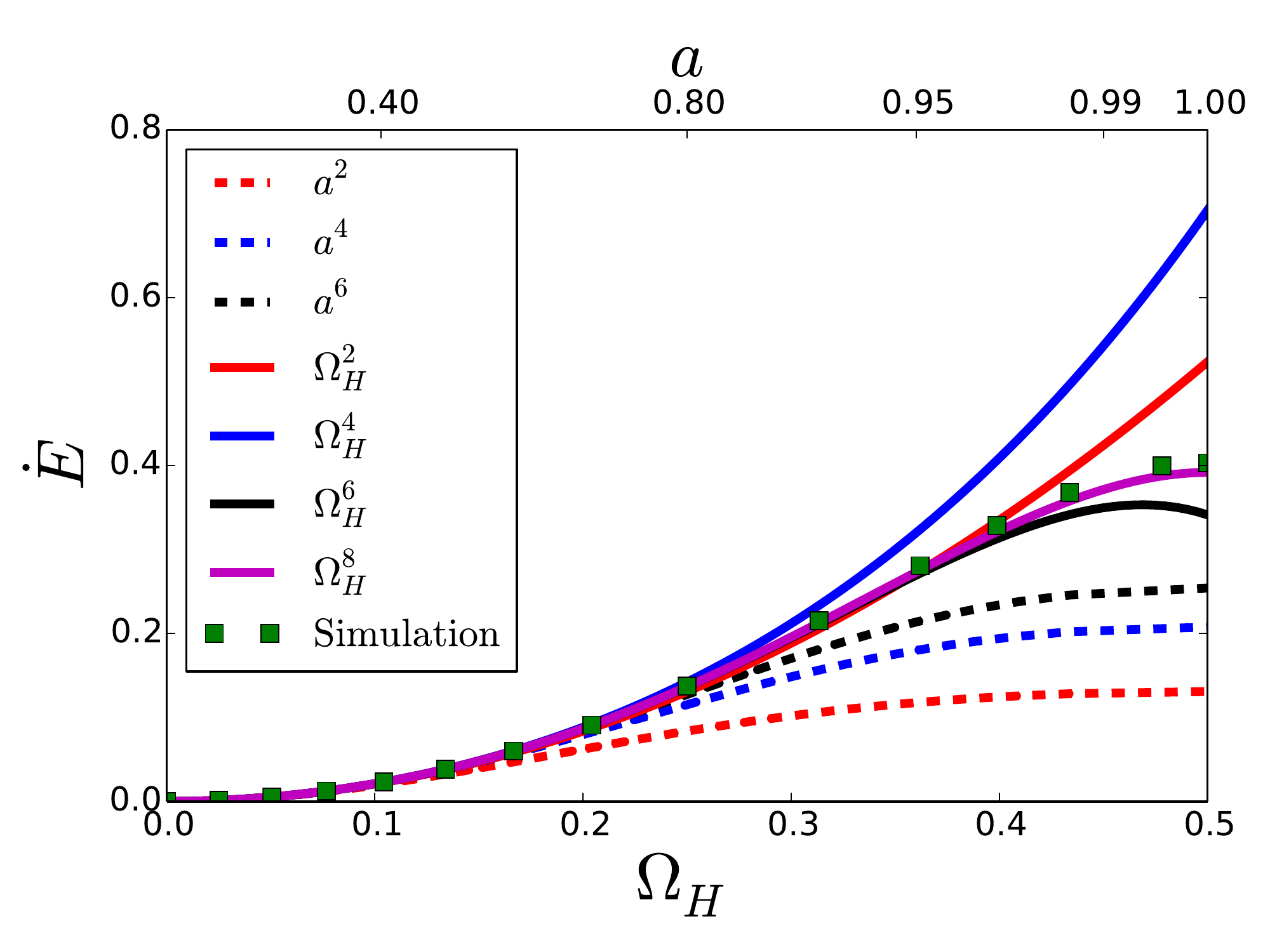}
\caption{The energy extraction rate $\dot E$ accurate to different orders.}
\label{fig:edot}
\end{figure}

The energy extraction rate defined by
$\dot E \equiv -2\pi\int_0^\pi \sqrt{-g} T^r_{\ t} d\theta = 2\pi \int I(A_\phi)\Omega(A_\phi) d A_\phi$, where $T^\mu_{\ \nu}$ is energy-momentum tensor of electromagnetic field, is calculated accurate to $O(a^6)$ as
\ben
\dot E
&=& 2\pi \int i(A_0) \omega(A_0) d A_0\nn\\
&=&\frac{\pi}{24}(a^2+0.587a^4+0.355a^6).
\een
Note that the 6th-order correction to the energy extraction rate only depends on $i_{1,3,5}$ and $\omega_{1,3,5}$ which are determined by $A_{0,2,4}$.
See Fig. \ref{fig:edot} for the energy extraction rate accurate to
different orders. For comparison, corresponding simulation results
from \citet{Tchekhovskoy2010} are also plotted. We find that the
analytic energy extraction rate of $O(a^6)$ better matches
simulation results compared to the original second-order BZ
solution, but still underestimates energy extraction rate by $\sim
30\%$ for extreme spins, which indicates that even higher orders
should be included. To improve the performance of analytic
solutions, one way is to expand them to higher order of spin. We
follow an alternative approach in this paper. We aims to find
perturbation expansion series of faster convergence.

\begin{figure}
\includegraphics[scale=0.42]{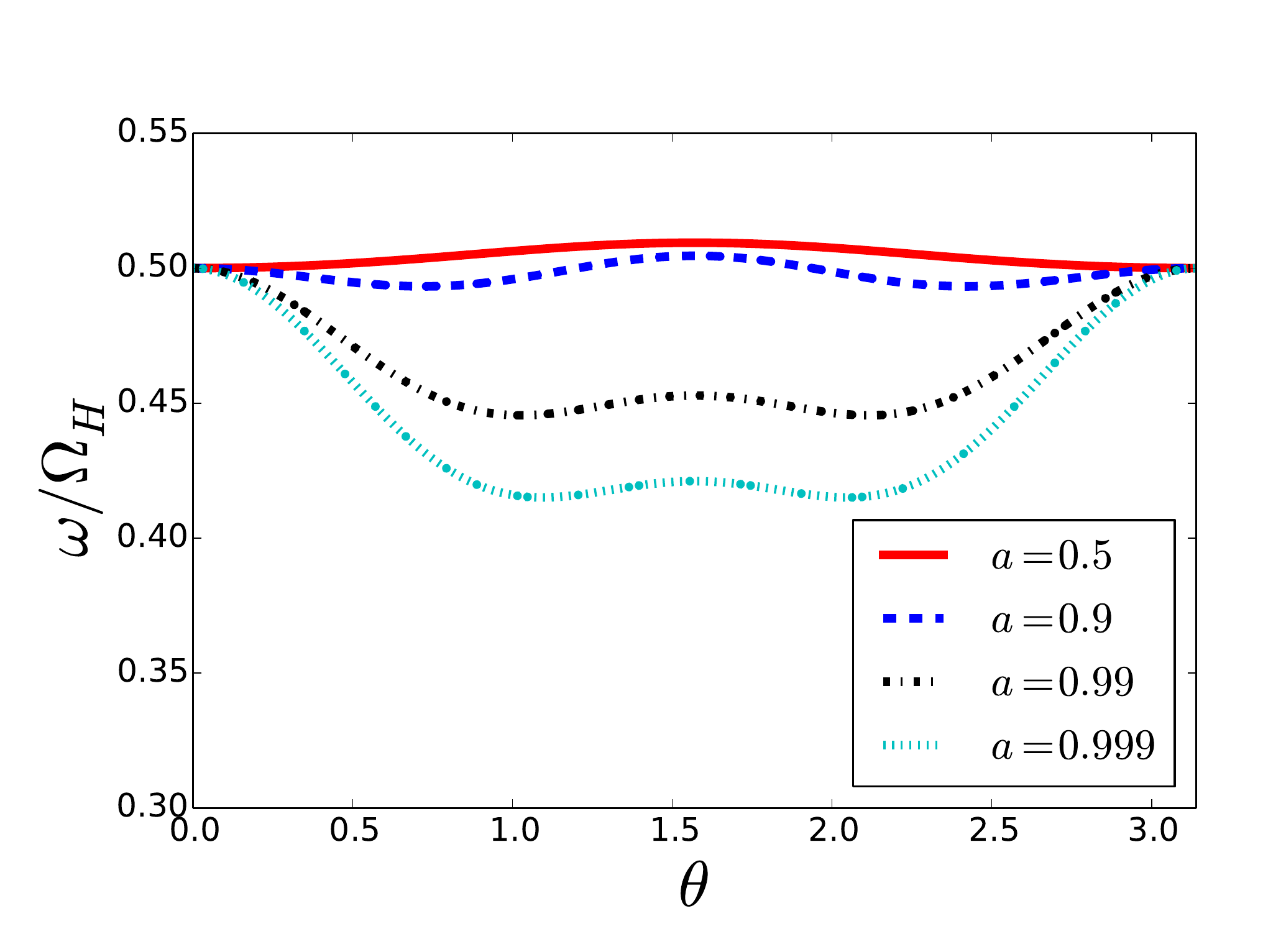}
\caption{Distribution of angular velocity of magnetic field lines $\omega$ accurate to $O(\WH^6)$.}
\label{fig:omega}
\end{figure}

Simulations and analytic works agree that, $\omega = \WH/2$ is a
good approximation, where $\WH = a/(2+2\sqrt{1-a^2})$ is angular
velocity of a Kerr BH. We expect a faster solution convergence in
terms of $\WH$ instead of $a$. Accurate to $O(\WH^6)$, we have \be
\omega = \frac{1}{2}\WH +  0.43\WH^3 \sst + \WH^5(-5.75\sin
^2\theta+2.48\sin ^4\theta), \ee which is shown in Fig.
\ref{fig:omega}. We see $\omega\simeq \WH/2$ for $a \lesssim 0.9$
and $\omega$ decrease abruptly for $a\gtrsim 0.9$. This turn-over
was seen in many numerical simulations and is analytically
confirmed here for the first time. In addition, take the extreme
spin case $a=0.999$ for example, we see the substructure that
$\omega$ decreases quickly then increases slowly with increasing
angle $\theta$ in the range of $0 \leq \theta \leq \pi/2$. The
substructure was also confirmed by previous simulations \citep{Tchekhovskoy2010}.

With $\omega$ expressed in terms of $\WH$, the energy extraction
rate is expressed in terms of $\WH$ as \be \dot E = \frac{2\pi}{3}
(\WH^2 + 1.38 \WH^4 -11.09 \WH^6). \ee There is a smart way to
extend $\dot E$ to even higher order making use of the relation
\be\label{smartbc}
\left(\frac{d\dot E}{d\WH}\right)_{a=1} =
\left(\frac{d\dot E}{da}\frac{da}{d\WH}\right)_{a=1} = 0, \ee  where we
have used the empirical constraint that $d\dot E/da$ is finite and
the relation $(da/d\WH)_{a=1}=0$. As a result,
eighth-order energy extraction rate $\dot E$ can be written as
\be
\dot E = \frac{2\pi}{3} (\WH^2 + 1.38 \WH^4 -11.09 \WH^6 + 6.26 \WH^8).
\ee
Note that the eighth-order correction here is in principle different
from the true correction that would be obtained in the usual way by solving
 the GS equation.  The energy extraction rate $\dot E$ accurate to
different orders are shown in Fig. \ref{fig:edot}. It turns out
that our analytic solution up to $O(\WH^8)$ perfectly reproduces
the result of simulations for the whole range of spins. Besides,
the commonly used lowest-order approximation $\dot E = (2\pi/3)
\WH^2$ overestimates the energy extraction rate by $\sim 30\%$ for
the extreme spin, because, it does not account for the abrupt
decrease of $\omega$ from $\WH/2$ for $a\gtrsim 0.9$.

\section{Discussion}
\label{sec: discussions}

We find that $i_n = \omega_n \sst$ for
$n\leq5$ for split-monopole magnetic field configuration.
Generally, we could prove the relation hold for all positive
integer $n$.  Take the $n=8$ case for example,
\ben
S_8
&=&
\sst\omega_1(\sst\omega_7)_{,\theta}+\sst\omega_3(\sst\omega_5)_{,\theta} \nn\\
&+&
\sst\omega_5(\sst\omega_3)_{,\theta}+\sst\omega_7(\sst\omega_1)_{,\theta}\nn\\
&-&(i_1 i_{7,\theta}+i_3 i_{5,\theta}+i_5 i_{3,\theta}+i_7 i_{1,\theta})+O(1/r),
\een
where we have explicitely listed all source terms of $O(1)$. According to the convergence requirement, the summation of these  $O(1)$ terms vanishes, so we have $i_7 = \omega_7\sst$. Similarly, we find $i_n = \omega_n \sst$ for all positive $n$. As a result, we have $I=\Omega \sst|_{r\rightarrow\infty}$. Note that both $I$ and $\Omega$ are functions of $A_\phi$, so $I/\Omega$ is also a function $A_{\phi}$. We denote it as $ I/\Omega \equiv f(A_\phi)$. This function can be readily determined at infinity. It is known that $I / \Omega |_{r\rightarrow\infty} = \sin^2\theta$ and $A_\phi|_{r\rightarrow\infty} = -\cos\theta$, so we have $f(-\cos\theta) = \sin^2\theta$ or $f(x) = 1-x^2$. 
Finally, we arrive at an exact constraint relation
\be\label{eq:exact}
I =\Omega (1-A_\phi^2),
\ee
which holds at all radii. Recently, \citet{Penna2015} shows that our constraint
relation (\ref{eq:exact}) is equivalent to the outgoing boundary condition at infinity.

\begin{figure}
\includegraphics[scale=0.45]{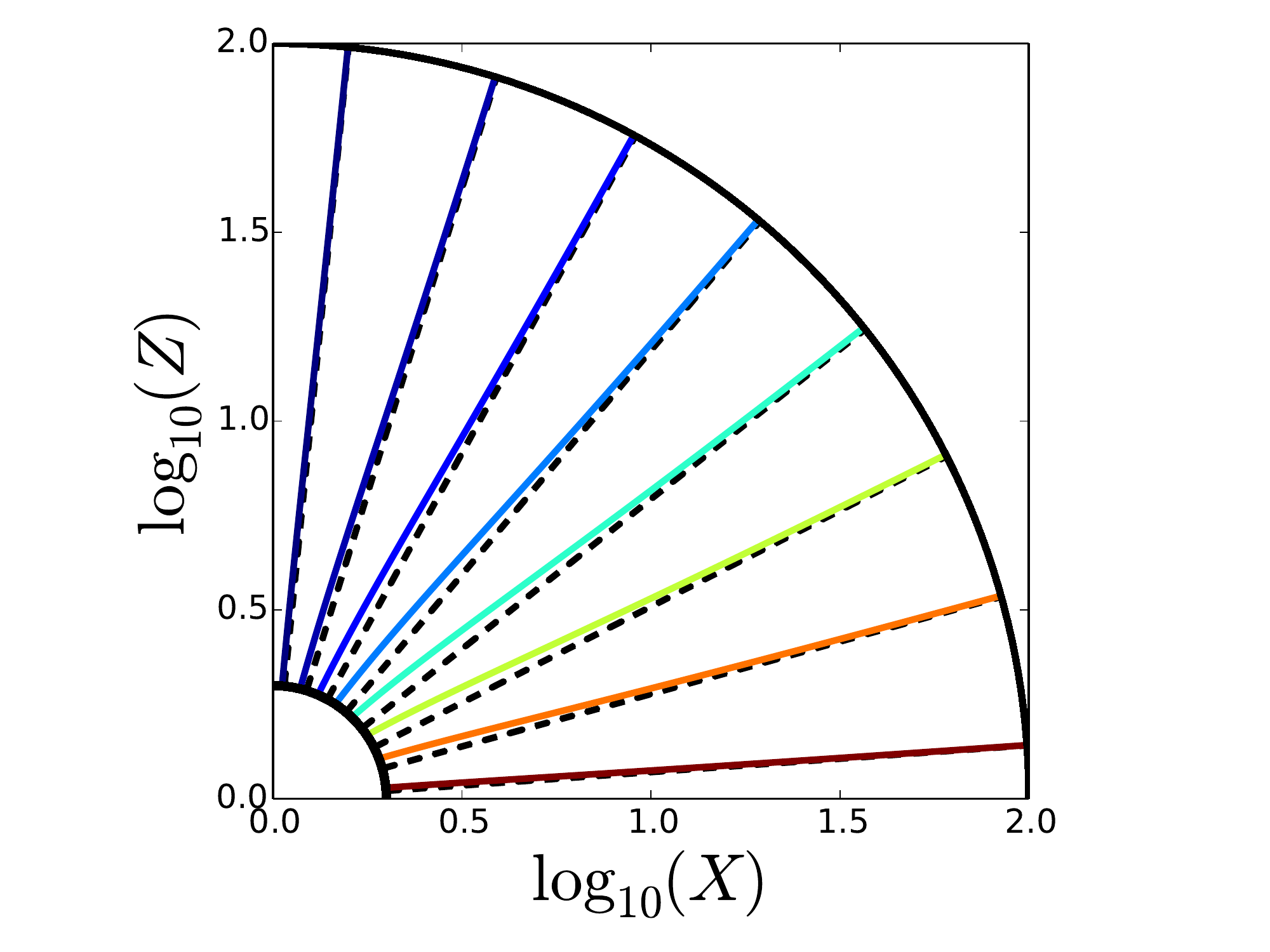}
\caption{Illustration of distortions of poloidal field lines induced by BH rotation, where solid lines are poloidal field lines in Kerr spacetime and dashed lines are corresponding field lines in Schwarzschild spacetime.}
\label{fig:poloidal}
\end{figure}

According to this constraint relation, we show that the
distortions of poloidal lines and variations of angular velocity
of magnetic field lines are in fact results of force balance of
magnetic field. We start with the explanation of $A_{\phi,\theta}$
variation with BH spin: BH rotation induces a toroidal component
of magnetic field $\mathbf{B_T}$, whose hoop stress exerts a
polar-directed force on poloidal field lines. In response,
poloidal field lines move towards the polar direction (see Fig.
\ref{fig:poloidal}), thus the fraction of magnetic flux increases
in the polar region and decreases in the equatorial region. In
other word, $|A_{\phi,\theta}|$ increases in the polar region and
decreases in  the equatorial region due to hoop stress of toroidal
field generated by BH rotation. Combing the constraint relation (\ref{eq:exact})
with the Znajek regularity condition, $\Omega$ on horizon $r=r_+$
can be expressed as \be \frac{\Omega}{\WH-\Omega} =
\left(\frac{\st
A_{\phi,\theta}}{1-A^2_\phi}\right)\left(\frac{2r_+}{r_+^2+a^2\cct}\right),
\ee so $\Omega$ is determined by distortions of poloidal field
lines and the size of BH horizon. For simplicity, we consider
$\Omega$ in two regions: the equatorial region $\theta\simeq\pi/2$
and the polar region $\theta\simeq 0$. For the former, \be
\left(\frac{\Omega}{\WH-\Omega} \simeq \frac{2
A_{\phi,\theta}}{r_+}\right)_{\theta\simeq\pi/2}.
\label{eq:forcebalance} \ee The right-hand side of the above
equation is proportional to $\mathbf{B_R/B_T}$, where the toroidal
field $\mathbf{B_T} \sim 1/r$ and radial field $\mathbf{B_R}\sim
A_{\phi,\theta}/r^2$. Decreasing $A_{\phi,\theta}$ will suppress
$\mathbf{B_R}$. To keep the force balance, $\Omega$ must decrease
accordingly to suppress hoop stress. Decreasing $r_+$ will
increase more $\mathbf{B_R}$ than $\mathbf{B_T}$. To keep the force balance,
$\Omega$ must increase accordingly. In real situation, both $r_+$
and $A_{\phi,\theta}$ decrease with increasing spin $a$, and
$\Omega/\WH$ is determined by the competition between the two. For
the polar region, we expect vanishing distortions of poloidal
field lines, due to vanishing toroidal field, then we have \be
\Omega|_{\theta=0} = \WH/2, \ee for any BH spin. Similar arguments
also apply to other field configurations, e.g., magnetic field
lines moving towards polar region and the turning over of
$\Omega/\WH$ were indeed seen in simulations of paraboloidal
magnetic field \citep{Tchekhovskoy2010}.

\section{Conclusions}
\label{sec: conclusions}

A stationary axisymmetric force-free jet
launched by the BZ mechanism is completely determined by three
function, $A_\phi$, $I(A_\phi)$ and $\Omega(A_\phi)$, which are
related by the GS equation (\ref{eq:GS}). We propose an
analytic approach to self-consistently determine the three
functions. As a specific application, we present a high-order
split monopole perturbation solution to the BZ mechanism. The
solution accurately describes the structures of BH magnetosphere
and the energy extraction rates of BZ mechanism for the whole
range of BH spins. The analytic approach also yields an exact
constraint relation, $I = \Omega (1-A^2_\phi)$, for the split
monopole magnetic field in the Kerr spacetime. The constraint
relation provides a useful tool for testing accuracy of numerical
simulations. Based on this constraint relation, we find that the
distortions of magnetic field lines and the variations of angular
velocity $\Omega$ of magnetic field lines are results of force
balance between the poloidal and toroidal field. \\

CY is grateful for the support by the National Natural Science Foundation
of China (Grant 11173057, 11373064, 11521303), Yunnan Natural Science
Foundation (Grant 2012FB187, 2014HB048). Part of the computation was performed
at the HPC Center, Yunnan Observatories, CAS, China.
This work made extensive use of the NASA Astrophysics Data System and
of the {\tt astro-ph} preprint archive at {\tt arXiv.org}.


\end{document}